\documentclass[conference]{IEEEtran}

\usepackage{cite}
\usepackage{amsmath,amssymb,amsfonts}
\usepackage{graphicx}
\usepackage{textcomp}
\usepackage{enumitem}
\usepackage{booktabs}
\usepackage{algorithm}
\usepackage{algorithmic}
\usepackage{seqsplit}
\usepackage{balance}
\usepackage{hyperref}
\usepackage{url}
\usepackage{xcolor}
\usepackage{comment}

\usepackage{scalerel} 
\usepackage{tikz} 
\usetikzlibrary{svg.path} \definecolor{orcidlogocol}{HTML}{A6CE39} \tikzset{ orcidlogo/.pic={ \fill[orcidlogocol] svg{M256,128c0,70.7-57.3,128-128,128C57.3,256,0,198.7,0,128C0,57.3,57.3,0,128,0C198.7,0,256,57.3,256,128z}; \fill[white] svg{M86.3,186.2H70.9V79.1h15.4v48.4V186.2z} svg{M108.9,79.1h41.6c39.6,0,57,28.3,57,53.6c0,27.5-21.5,53.6-56.8,53.6h-41.8V79.1z M124.3,172.4h24.5c34.9,0,42.9-26.5,42.9-39.7c0-21.5-13.7-39.7-43.7-39.7h-23.7V172.4z} svg{M88.7,56.8c0,5.5-4.5,10.1-10.1,10.1c-5.6,0-10.1-4.6-10.1-10.1c0-5.6,4.5-10.1,10.1-10.1C84.2,46.7,88.7,51.3,88.7,56.8z}; } } 

\newcommand\orcid[1]{\href{https://orcid.org/#1}{\mbox{\scalerel*{
\begin{tikzpicture}[yscale=-1,transform shape]
\pic{orcidlogo};
\end{tikzpicture}
}{|}}}}

\newcommand{\simf}{\operatorname{sim}}

\def\BibTeX{{\rm B\kern-.05em{\sc i\kern-.025em b}\kern-.08em
    T\kern-.1667em\lower.7ex\hbox{E}\kern-.125emX}}

\usepackage{eso-pic}

\AddToShipoutPictureFG{%
  \AtPageLowerLeft{%
    \makebox[\paperwidth][c]{%
      \raisebox{20pt}{%
        \footnotesize
        Distribution Statement ``A'' (Approved for Public Release, Distribution Unlimited)%
      }%
    }%
  }%
}
\begin{document}
\IEEEoverridecommandlockouts


\title{An Interpretable Approach to Money Laundering Detection in Transaction Graphs using Pass-Through Templates}

\author{\IEEEauthorblockN{ Paolo Climaco  \orcid{0000-0003-1280-4930} \IEEEcompsocitemizethanks{\IEEEcompsocthanksitem This research was, in part, funded by the US Government.  The views and conclusions contained in this document are those of the author and should not be interpreted as representing the official policies, either expressed or implied, of the US Government.}}
\IEEEauthorblockA{\textit{Department of Mathematics} \\
\textit{University of California, Los Angeles}\\
Los Angeles, CA, USA\\
climaco@math.ucla.edu}

}

\maketitle

\begin{abstract}
\emph{Layering} is a key stage of the money laundering process in which assets are moved through intermediate entities to obscure their origin. We model this movement as a transaction graph, a labeled directed multigraph where nodes represent entities, such as individuals or organizations, and edges represent transactions between them. This work addresses the detection of layering patterns in transaction graphs. We define \emph{pass-through} templates, a class of transaction graphs whose structure is indicative of a specific layering pattern in which entities receive funds and rapidly forward them. We formulate detecting instances of these templates within a larger transaction graph as a maximum-cardinality, maximum-weight matching problem. The resulting method uses Edmonds' blossom algorithm, which provably finds an optimal solution to this problem. Here we present a case study on publicly available Ethereum transaction data. We identify 283 pass-through template instances, chained through shared address nodes into a densely interconnected network that moves hundreds of millions of dollars in stablecoin value. We find that six addresses in the identified network are sanctioned by the U.S. Department of the Treasury for associations with criminal organizations, providing evidence of ties between the network and sanctioned criminal entities.
\end{abstract}


\section{Introduction}
Money laundering conceals the link between illicit funds and their source through three stages: placement, in which the funds enter a financial system; layering, in which they are moved through transactions to obscure their origin; and integration, in which they are made to appear legitimate, for example by investing in real estate or businesses~\cite{fincen_msb_prevention}. We model the movement of funds as a labeled directed multigraph, which we call a transaction graph. Its nodes represent entities, such as individuals or organizations, and its edges represent the transactions between them labeled by asset, amount, and time. This work focuses on detecting layering patterns in transaction graphs.

The rapid pass-through of value is a common layering pattern in which an entity receives funds and transfers them onward shortly afterward in the same or a similar amount~\cite{fincen2010tbml,federalreserve2001scrutiny}. We formalize this pattern as \emph{pass-through templates}: a class of transaction graphs defined by explicit criteria, in which a designated node receives funds and rapidly forwards them. We systematically search large transaction graphs for instances of these templates. Detection is performed at the node level by determining whether the one-hop neighborhood of each node matches a pass-through template. We formulate this as a maximum-cardinality, maximum-weight matching problem between incoming and outgoing transfers, with compatibility determined by asset, timing, and amount, and solve it using the blossom algorithm of Edmonds~\cite{Edmonds1965MaximumMA,Banerjee2008}.

Existing graph-based approaches to detecting laundering-related activity typically rely on labeled training data~\cite{weber2019anti,bellei2024shape,song2024identifying}, flag anomalously dense subgraphs~\cite{hooi2016fraudar,liu2017holoscope,shin2016mzoom}, or assume a fixed template topology for subgraph matching~\cite{ullmann1976algorithm,Moorman2021Subgraph}. Our method requires no learned model or labeled examples and does not assume a fixed topology. Each match represents an explicit pass-through pattern associated with layering and is interpretable through clearly defined criteria on graph structure, asset, timing, and amount.

We test our approach in a case study on the Ethereum blockchain.
We build a pipeline that extracts transactions from raw, publicly available data and assembles a graph where nodes represent addresses and edges represent transactions of digital assets, such as stablecoins and ETH. We construct a transaction graph by expanding outward from a seed address. This enables a focused analysis of a subnetwork rather than the full Ethereum network, reducing the computational burden. The seed address is selected from a list of sanctioned addresses publicly released by the U.S. Department of the Treasury's Office of Foreign Assets Control (OFAC)~\cite{ofac_sdn_list}. 

Our analysis identifies 283 pass-through template instances that chain together through shared nodes into a single 1,690-address layering network: 98.3\% of identified nodes fall into one connected component, consistent with a coordinated operation rather than isolated instances of pass-through patterns. We find that the OFAC-sanctioned seed address is in the largest component of the layering network alongside five additional OFAC-sanctioned addresses. Over eight months, the network moves more than \$260 million through matched pass-through transfers. Overall, its nodes receive roughly \$410 million in stablecoin value from outside sources.

The remainder of this paper is organized as follows. Section~\ref{sec:related} reviews related work; Section~\ref{sec:methodology} formalizes transaction graphs, pass-through templates, and the matching procedure; Section~\ref{sec:data} describes the Ethereum data-construction pipeline and experimental setup; and Section~\ref{sec:results} analyzes the structure and value flow of the resulting layering network.

\section{Related Work}
\label{sec:related}
Machine-learning methods are widely applied to transaction graphs for anti-money-laundering detection. Weber et al.~\cite{weber2019anti} formulate illicit-transaction detection on the Bitcoin blockchain as a classification problem and investigate graph convolutional networks and standard machine-learning classifiers. More recent work has shifted from individual transactions to larger transaction structures. Bellei et al.~\cite{bellei2024shape} introduce the Elliptic2 dataset and formulate cryptocurrency money-laundering detection as a subgraph classification problem, where suspicious transaction subgraphs are identified from labeled examples. Song et al.~\cite{song2024identifying} further develop this approach by tracking the sources and destinations of funds and using this for subgraph classification and discovery. These methods learn statistical representations of illicit activity from labeled data, whereas we seek to identify an explicitly defined transaction behavior without requiring labeled training examples.

A related class of methods detects fraudulent activity through anomalously dense subgraphs. FRAUDAR~\cite{hooi2016fraudar} identifies suspicious dense activity in bipartite graphs using a scoring function robust to adversarial camouflage edges. HoloScope~\cite{liu2017holoscope} and M-Zoom~\cite{shin2016mzoom} extend dense-subgraph detection by incorporating additional edge information: HoloScope uses temporal activity patterns and rating scores, while M-Zoom considers multiple transaction attributes simultaneously. 

Flow-based methods directly model the movement of funds through transaction networks. FlowScope~\cite{li2020flowscope} identifies anomalous source-to-destination flows through multipartite graphs, favoring high transaction volume while penalizing imbalances at intermediary accounts. CubeFlow~\cite{sun2021cubeflow} extends flow-based detection to attributed transaction data. It models transactions as two coupled tensors and identifies suspicious transfer chains through intermediary accounts using a multi-attribute flow metric. In contrast, we detect pass-through behavior locally by matching incoming and outgoing transactions within the one-hop neighborhood of each node under asset, temporal, and amount-similarity constraints, formulated as a maximum-cardinality, maximum-weight matching problem. Our analysis applies to general transaction networks, not only multipartite graph structures.

Our formulation is also related to subgraph matching, where a smaller template graph is sought within a larger graph. Exact subgraph isomorphism methods~\cite{ullmann1976algorithm, Moorman2021Subgraph} search for mappings that preserve the topology of a fixed template. Inexact matching methods relax these constraints by permitting structural or label mismatches under an application-dependent penalty~\cite{Khan2013NeMa}. In contrast, pass-through templates do not prescribe a single fixed topology: the number of incoming and outgoing transactions may vary across nodes. We therefore define a class of admissible templates through transaction constraints and exploit their structure to reduce matching to an optimization problem.

\section{Problem Setting and Methodology}
\label{sec:methodology}
This section defines transaction graphs and pass-through templates formally, then formulates template detection as a maximum-cardinality, maximum-weight matching problem.
\subsection{Transaction Graphs}
\label{sec:blockchain-graph}
Transaction graphs model the transfer of assets between entities. We define a transaction graph as a directed edge-labeled multigraph of the form
\begin{equation}
G = (V, E, \lambda_E),
\label{eq:worldgraph}
\end{equation}
where $V$ is the finite node set of observed entities, such as individuals or organizations, and $E$ is the finite multiset of directed transfer edges. $\lambda_E$ is a map that labels each edge with the transacted asset, amount, and time. We define the maps $$\mathrm{src}: E \rightarrow V \quad \textrm{and}  \quad \mathrm{dst}: E \rightarrow V,$$  determining the source and destination of each edge, respectively. An edge $e \in E$ is outgoing for $\mathrm{src}(e) \in V$ and incoming for $\mathrm{dst}(e)\in V$. 

Let $\mathcal{A}$ denote the set of all assets, e.g., state-backed currencies, crypto assets or tradable commodities. The edge-label map $\lambda_E : E \to\mathcal{A} \times \mathbb{R}_{>0} \times \mathbb{N}$ assigns each edge a tuple of the form
 \begin{equation}
 \lambda_E(e) = \bigl(a_e, p_e, \tau_e\bigr) \in \mathcal{A} \times \mathbb{R}_{>0} \times \mathbb{N},
 \label{eq:edgelabel}
 \end{equation}
 where $a_e$ is the asset, $p_e$ is the transferred amount, and $\tau_e$ is the time-stamp indicating when the transaction occurred. Since $G$ is a multigraph, two distinct edges $e_1 \neq e_2$ may share the same source and destination while representing distinct transactions. That is, $\mathrm{src}(e_1) = \mathrm{src}(e_2)$, $\mathrm{dst}(e_1) = \mathrm{dst}(e_2)$, and $\lambda_E(e_1) \neq \lambda_E(e_2)$. Every transaction graph we consider uses the same edge-label map, thus we drop $\lambda_E$ from the notation when unambiguous and simply write $G = (V, E)$. We do not consider node labels here, but the framework extends naturally to incorporate them if needed.

\subsection{Pass-Through Templates}
\label{sec:qualifying}
A \emph{pass-through template} is a transaction graph that captures a layering pattern in which a designated node $x_c$ receives funds and quickly forwards the same or a similar amount onward. Formally, we denote it by
\[
G_t(x_c) = (V_t(x_c), E_t(x_c), \lambda_E).
\]
Here, $x_c \in V_t(x_c)$ is the only node with both incoming and outgoing edges. The edge multiset $E_t(x_c)$ consists of paired incoming and outgoing transaction edges, representing funds received by $x_c$ and then passed onward. That is, 
\[
\begin{aligned}
E_t(x_c) := &\bigcup_{i=1}^m \{e_i^{\mathrm{in}}, e_i^{\mathrm{out}}\}, \quad m \geq 1, \\
&\text{where } \mathrm{dst}(e_i^{\mathrm{in}}) = x_c,\ \ \mathrm{src}(e_i^{\mathrm{out}}) = x_c, \\
&\mathrm{src}(e_i^{\mathrm{in}}),\ \mathrm{dst}(e_i^{\mathrm{out}}) \in V_t(x_c) \setminus \{x_c\}.
\end{aligned}
\]
$e_i^{\mathrm{in}}$ is an incoming edge to $x_c$ and $e_i^{\mathrm{out}}$ is the paired outgoing edge for $i=1,\dots,m$. Each pair $\{e_i^{\mathrm{in}},e_i^{\mathrm{out}}\}$ represents a single pass-through, where $x_c$ receives and then forwards value.
All incoming edges must be distinct and
all outgoing edges must be distinct, so that each transaction belongs to exactly one pair of incoming and outgoing edges.

To be consistent with layering, each pair must satisfy a set of qualifying criteria: the outgoing transaction must happen shortly after the incoming one, in the same asset, and for a similar amount.
Consider
\begin{equation*}
\lambda_E\bigl(e_i^{\mathrm{in}}\bigr) = \bigl(a_{\mathrm{in},i}, p_{\mathrm{in},i}, \tau_{\mathrm{in},i}\bigr),
\end{equation*}
\begin{equation*}
\lambda_E\bigl(e_i^{\mathrm{out}}\bigr) = \bigl(a_{\mathrm{out},i}, p_{\mathrm{out},i}, \tau_{\mathrm{out},i}\bigr).
\end{equation*}
Then the pair $\{e_i^{\mathrm{in}}, e_i^{\mathrm{out}}\}$ must satisfy the qualifying criteria outlined below.

\begin{enumerate}[label=\Roman*.]

\item \textbf{Same asset:} The paired edges must have the same asset label, that is, we require $a_{\mathrm{in},i} = a_{\mathrm{out},i}$.

\item \textbf{Pass-through behavior:} Value must flow through $x_{c}$ within a given time window. The transaction times must satisfy
\begin{equation}
0 < \tau_{\mathrm{out},i} - \tau_{\mathrm{in},i} \le \bar{T},
\label{eq:temporal}
\end{equation}
where $\bar{T}$ is a user-specified maximum time gap. 

\item \textbf{Amount similarity:} We require the in amount, $p_{\mathrm{in},i}$ and the out amount $p_{\mathrm{out},i}$ to match or be similar. We use the following similarity function
\begin{equation}
\simf(p_{\mathrm{in},i}, p_{\mathrm{out},i}) = 1 - \frac{|p_{\mathrm{out},i} - p_{\mathrm{in},i}|}{\max(p_{\mathrm{in},i}, p_{\mathrm{out},i})}.
\label{eq:similarity}
\end{equation}
 The pair $\{e_i^{\mathrm{in}}, e_i^{\mathrm{out}}\}$ qualifies only if $\simf(p_{\mathrm{in},i}, p_{\mathrm{out},i}) \ge \epsilon_{\mathrm{sim}}$, where $\epsilon_{\mathrm{sim}}$ is a user-specified similarity threshold.

\item \textbf{Minimum transacted value:} We require the amount associated with each edge to exceed a given threshold, i.e, $\mathrm{min}\{p_{\mathrm{in},i}, p_{\mathrm{out},i}\} \ge p_{\mathrm{min}}$.  
\end{enumerate}
We call a pair of edges satisfying Conditions~I--IV a \emph{qualifying pair}.
In addition to the conditions above, we require that the number of qualifying pairs $m=\frac{|E_{t}(x_c)|}{2}$, associated with a candidate pass-through node $x_{c}$, is at least as large as a given threshold $m_{\min} \geq 1$. If this condition is not met, $G_t(x_{c})$ is not considered a pass-through template.


Pass-through templates only need to respect the requirements defined above and may therefore have different topologies. 
\subsection{Methodology}
\label{sec:optimization}
We detect pass-through templates by testing, for each node in a transaction graph, whether its one-hop neighborhood matches a pass-through template structure. Consider a transaction graph $G =(V,E)$. For a fixed candidate pass-through node $x_{c} \in V$, we seek the pass-through template through $x_{c}$ with the largest number $m$ of qualifying pairs. This maximum-cardinality template captures the full extent of pass-through behavior at $x_{c}$. Denote the set of incoming edges and outgoing edges at $x_{c}$ as
\begin{equation}
  \label{eq:in-out-edges-sets}
\begin{aligned}
\mathcal{I}(x_{c}) &:= \{e \in E : \mathrm{dst}(e) = x_c,\ \mathrm{src}(e) \neq x_c\}, \\
\mathcal{O}(x_{c}) &:= \{e \in E : \mathrm{src}(e) = x_c,\ \mathrm{dst}(e) \neq x_c\},
\end{aligned}
\end{equation}
 respectively. For any candidate pair $\bigl(e_i^{\mathrm{in}}, e_j^{\mathrm{out}}\bigr) \in \mathcal{I}(x_{c}) \times \mathcal{O}(x_{c})$, write
\begin{equation*}
\begin{aligned}
\lambda_E\bigl(e_i^{\mathrm{in}}\bigr)& =
\bigl(a_{\mathrm{in},i}, p_{\mathrm{in},i}, \tau_{\mathrm{in},i}\bigr), \\
\lambda_E\bigl(e_j^{\mathrm{out}}\bigr)& =
\bigl(a_{\mathrm{out},j}, p_{\mathrm{out},j}, \tau_{\mathrm{out},j}\bigr).
\end{aligned}
\end{equation*}
We then define the compatibility set of all qualifying pairs by
\begin{equation}
\begin{aligned}
\mathcal{Q}(x_{c}) = \Bigl\{ &\bigl(e_i^{\mathrm{in}}, e_j^{\mathrm{out}}\bigr) :
e_i^{\mathrm{in}} \in \mathcal{I}(x_{c}),
e_j^{\mathrm{out}} \in \mathcal{O}(x_{c}), \\
& a_{\mathrm{in},i} = a_{\mathrm{out},j}, \\
& 0 < \tau_{\mathrm{out},j} - \tau_{\mathrm{in},i} \le \bar{T}, \\
& \simf(p_{\mathrm{in},i}, p_{\mathrm{out},j}) \ge \epsilon_{\mathrm{sim}}, \\
& \min\{p_{\mathrm{in},i}, p_{\mathrm{out},j}\} \ge p_{\min} \Bigr\}.
\end{aligned}
\label{eq:compatibility-set}
\end{equation}
Thus, $\mathcal{Q}(x_c) \subseteq \mathcal{I}(x_{c})\times \mathcal{O}(x_{c}) $ consists of the pairs satisfying Conditions~I--IV of Section~\ref{sec:qualifying}. A pass-through template through $x_{c}$ is therefore determined by a subset $S \subseteq \mathcal{Q}(x_{c})$ such that no incoming edge in $\mathcal{I}(x_{c})$ and no outgoing edge in $\mathcal{O}(x_{c})$ appears in more than one element of $S$. Thus, template matching reduces to finding such a subset $S$ of maximum cardinality.

Maximizing cardinality alone need not identify a unique match: several feasible selections may attain the same optimal cardinality while representing distinct value flows. For example, suppose two incoming transfers have amounts 100 and 101, and two outgoing transfers also have amounts 100 and 101. A 1\% dissimilarity tolerance (Condition~III, $\epsilon_{\mathrm{sim}}=0.99$) would allow both pairings, so two distinct selections of cardinality two are feasible, one pairing the equal amounts exactly and the other not. Resolving this ambiguity helps us select the value flow most consistent with canonical pass-through patterns.

We resolve such ties \emph{lexicographically}~\cite{Isermann1982}, the same way words are ordered in a dictionary: the first letter decides the order, and later letters matter only when the earlier ones are equal. Cardinality is optimized first, and total amount similarity is used only to rank the selections that tie for the best cardinality.
For each $(e_i^{\mathrm{in}}, e_j^{\mathrm{out}}) \in \mathcal{Q}(x_{c})$, let $z_{ij} \in \{0,1\}$ indicate whether the pair is selected. We write $(i,j)\in\mathcal{Q}(x_c)$ as shorthand for $(e_i^{\mathrm{in}}, e_j^{\mathrm{out}}) \in \mathcal{Q}(x_c)$.  Formally, we solve
\begin{align}
\operatorname*{lex\,max}_{z_{ij} \in \{0,1\}} \ \ &
\biggl(\, \sum_{(i,j)\in \mathcal{Q}(x_{c})} \! z_{ij},\ \
\sum_{(i,j)\in \mathcal{Q}(x_{c})} \! \simf(p_{\mathrm{in},i}, p_{\mathrm{out},j})\, z_{ij} \biggr)
\label{eq:template-lexicographic} \\
\text{s.t.} \quad
& \sum_{j \,:\, (i,j)\in \mathcal{Q}(x_{c})} \! z_{ij} \le 1
\quad \forall\, e_i^{\mathrm{in}} \in \mathcal{I}(x_{c}), \notag \\
& \sum_{i \,:\, (i,j)\in \mathcal{Q}(x_{c})} \! z_{ij} \le 1
\quad \forall\, e_j^{\mathrm{out}} \in \mathcal{O}(x_{c}), \notag
\end{align}
where $\operatorname{lex\,max}$ ranks candidate solutions first by the first coordinate of the objective, and only among those tied on it, by the second. The first coordinate, $\sum_{(i,j)} z_{ij}$, counts how many pairs are selected. The second, $\sum_{(i,j)} \simf(p_{\mathrm{in},i}, p_{\mathrm{out},j})\, z_{ij}$, adds up the amount-similarity score of every selected pair. The two constraints ensure that no incoming or outgoing edge is used more than once.

Let $\{z_{ij}^\star\}$ be an optimal solution of \eqref{eq:template-lexicographic}. It defines the set of matched qualifying pairs
\begin{equation}
\label{eq:matched-qualifying-pairs}
M^\star(x_c):=
\{(e_i^{\mathrm{in}},e_j^{\mathrm{out}})\in\mathcal{Q}(x_c): z_{ij}^\star=1\}.
\end{equation}
Let us define the matched template edges as
\begin{equation}
  \label{eq:final-matched-template-edges}
E_t^\star(x_c)=
\bigcup_{(e^{\mathrm{in}},e^{\mathrm{out}})\in M^\star(x_c)} \hspace{-0.5cm}
\{e^{\mathrm{in}},e^{\mathrm{out}}\},
\end{equation}
and the matched template vertices as
\[
V_t^\star(x_c) := \{v \in V : \exists\, e \in E_t^\star(x_c),\ v \in \{\mathrm{src}(e), \mathrm{dst}(e)\}\}.
\]
Then, the final matched pass-through template through \(x_c\) is
\begin{equation}
  \label{eq:final-matched-template-graph_1}
  G_t^\star(x_c)=\bigl(V_t^\star(x_c),E_t^\star(x_c),\lambda_E\bigr).
\end{equation}
Note that $G_t^\star(x_c)$ is a subgraph of the transaction graph $G$.
\begin{algorithm}[t]
\caption{Pass-through template matching}
\label{alg:detect}
\begin{algorithmic}[1]
\REQUIRE Transaction graph $G=(V,E,\lambda_E)$ and thresholds $\bar{T}$, $\epsilon_{\mathrm{sim}}$, $p_{\min}$, $m_{\min}$
\ENSURE Matched templates $\mathcal{R} = \{G_t^\star(x_c)\}$
\STATE $\mathcal{R}\gets\varnothing$
\FOR{each $x_c\in V$}
  \STATE Define the set of incoming edges $\mathcal{I}(x_c)$ and outgoing edges $\mathcal{O}(x_c)$ at $x_c$ as in \eqref{eq:in-out-edges-sets}
  \IF{$|\mathcal{I}(x_c)| \ge m_{\min}$ and $|\mathcal{O}(x_c)| \ge m_{\min}$}
    \STATE form $\mathcal{Q}(x_c)$ as in \eqref{eq:compatibility-set} with $\bar{T}$, $\epsilon_{\mathrm{sim}}$, $p_{\min}$.
    \STATE build the weighted bipartite graph on $\mathcal{I}(x_c)\cup\mathcal{O}(x_c)$ with edge set $\mathcal{Q}(x_c)$
    \STATE compute $M^\star(x_c)$ with \texttt{\small nx.max\_weight\_matching} and \texttt{\small maxcardinality=True}
    \IF{$|M^\star(x_c)| \ge m_{\min}$}
      \STATE construct $G_t^\star(x_c)$ as in \eqref{eq:final-matched-template-graph_1}
      \STATE $\mathcal{R}\gets\mathcal{R}\cup\{G_t^\star(x_c)\}$
    \ENDIF
  \ENDIF
\ENDFOR
\STATE \textbf{return} $\mathcal{R}$
\end{algorithmic}
\end{algorithm}
We solve the lexicographic optimization problem in \eqref{eq:template-lexicographic} with the \texttt{nx.max\_weight\_matching} function from the NetworkX Python package~\cite{SciPyProceedings_11}. Run with the option \texttt{maxcardinality=True}, it maximizes cardinality first and breaks ties by total similarity weight in a single pass. This method implements Edmonds' blossom algorithm~\cite{Edmonds1965MaximumMA}, following~\cite{Galil1986,Banerjee2008}, which provably returns this lexicographic optimum for integer-valued weights. Real-valued weights are not an issue: the resulting matching is optimal up to floating-point rounding error, which is marginal in this work. Exact optimality can be guaranteed by scaling weights to integers.

Each per-node instance of the NetworkX matching function runs on the bipartite graph over $\mathcal{I}(x_c)$ and $\mathcal{O}(x_c)$ 
and takes $O\bigl((|\mathcal{I}(x_c)|+|\mathcal{O}(x_c)|)^3\bigr)$ time in floating point operations~\cite{Galil1986}. Let $D_{\max}:= \max_{x_c \in V} \bigl(|\mathcal{I}(x_c)|+|\mathcal{O}(x_c)|\bigr)$. Each instance costs at most $O(D_{\max}^3)$, so the total cost across all $|V|$ nodes is $O(D_{\max}^3 |V|)$: linear in the number of nodes once degree is bounded.

The matching procedure is summarized in Algorithm~\ref{alg:detect}. Note that the implemented approach is interpretable: every edge pair in a returned match satisfies Conditions~I--IV exactly, so the reason behind each matched instance is transparent and traceable to those explicit criteria.

\section{Data and Experimental Setting}
\label{sec:data}
This section describes the construction of the analyzed Ethereum transaction graph and the threshold parameters used to apply Algorithm~\ref{alg:detect} to it.

\subsection{Building the Ethereum Transaction Graph}
\label{sec:building-the-ethereum-transaction-graph}
The Ethereum blockchain is a public, distributed ledger that records transactions and state changes associated with Ethereum addresses. We obtain the raw data from the AWS Public Blockchain Data program~\cite{aws_public_blockchain}, which hosts a full public copy of the ledger. We instantiate the transaction graph of Section~\ref{sec:blockchain-graph} with Ethereum addresses as nodes and Ethereum transactions as edges. Each transaction can involve the transfer of assets, such as ETH or stablecoins like USDT and USDC \cite{watsky2024stablecoins}. 

The Ethereum network is massive: it involves more than 300 million addresses and more than 4.5 million daily token-transfers, at the time of writing. Due to computational constraints, we restrict our analysis in two ways: to a fixed set of major assets (native and wrapped ETH and leading stablecoins, including USDT, USDC, and DAI), and to a subnetwork around a seed address rather than the full Ethereum network. Focusing on stablecoins is a reasonable choice, since ``illicit actors using digital assets are reported to prefer stablecoins due to the relative stability compared to other digital assets"~\cite{treasury2026nmlra}.

\begin{figure}[t]
\centering
\includegraphics[width=\linewidth]{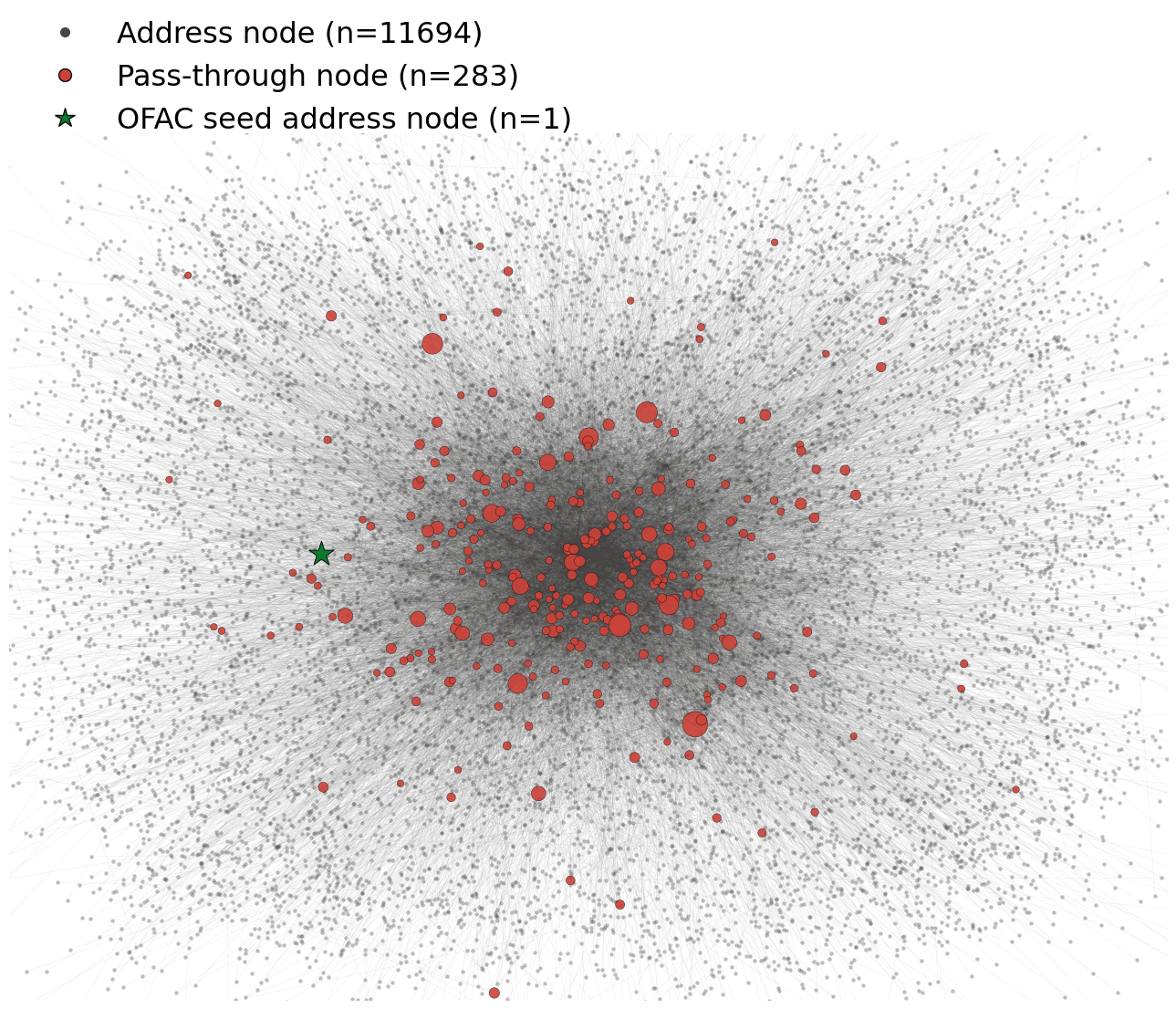}
\caption{Visualization of the Ethereum network analyzed in this work. It consists of 11{,}978 nodes and 49{,}921 edges (faint grey
structure). The 283 flagged pass-through nodes are drawn as larger red markers. The larger the markers, the more pass-through activity the associated nodes exhibit. The single OFAC seed address is marked with a green star.}
\label{fig:fullgraph}
\end{figure}

Specifically, we construct a subnetwork centered on a seed address drawn from the list of addresses sanctioned by the U.S. Department of the Treasury's Office of Foreign Assets Control (OFAC)~\cite{ofac_sdn_list}. No other information from OFAC is used to build the graph. Construction proceeds in two stages over the block-time window from 2022-08-06 to 2023-04-01. 
First, starting from the seed, we grow an address backbone through a breadth-first hop expansion: at each of $k=4$ hops, we identify transactions touching the current frontier and admit only those value-flow edges that connect back to an address already reached in an earlier hop, then add their new endpoints to the frontier of the next hop. Second, for every backbone address (excluding boundary addresses, discussed below), we separately record its complete inbound and outbound transfer activity over the same window and asset set, adding any previously unseen counterparty as a node. 
The resulting subnetwork contains 11{,}978 nodes and 49{,}921 edges. We denote this network as $G=(V,E)$. Table~\ref{tab:graph-composition} summarizes the size and edge composition of the network and Fig.~\ref{fig:fullgraph} illustrates the analyzed graph. 
\begin{table}[t]
\centering
\caption{Composition of the analyzed Ethereum transaction graph \(G=(V,E)\). The table reports the total number of address nodes and transaction edges, together with the number of edges associated with each transacted asset, comprising stablecoins and Ether. See Appendix~\ref{app:stablecoins} for additional information on these assets.}
\label{tab:graph-composition}
\setlength{\tabcolsep}{6pt}
\renewcommand{\arraystretch}{0.95}
\begin{tabular}{@{}lr@{}}
\toprule
\textbf{Quantity} & \textbf{Count} \\
\midrule
Nodes \(|V|\) & 11{,}978 \\
Edges \(|E|\) & 49{,}921 \\
\midrule
USDT edges & 44{,}227 \\
USDC edges & 4{,}616 \\
ETH edges & 703 \\
BUSD edges & 153 \\
WETH edges & 144 \\
DAI edges & 78 \\
\bottomrule
\end{tabular}
\end{table}
Each node represents an Ethereum address and stores attributes used later in the analysis, most importantly whether the address is treated as an externally owned account (EOA) or classified as a smart contract. Each edge represents an observed value transfer between two addresses and records the transfer time, the asset involved (e.g., the type of stablecoin), and the transferred amount.

To keep the expansion tractable, at extraction time we exclude 144 addresses from further frontier expansion: addresses whose transaction count exceeds a fixed threshold (600), since expanding through them would cause an unmanageable increase in tracked nodes, and addresses classified as smart-contract infrastructure or known high-volume entities (e.g., major exchange wallets), since they represent known infrastructure that adds little new information. We call the union of these addresses the boundary nodes $H \subset V$. The transaction activity of boundary nodes is still relevant to money laundering detection, but we do not expand the frontier through them. They remain in the extracted subnetwork as nodes and edge endpoints, and they may still gain edges when they appear as counterparties in transfers involving tracked addresses. Separately, we exclude known token-contract addresses, such as the USDT and USDC stablecoin contracts, from the graph entirely, since they represent protocol infrastructure rather than transacting entities and are not treated as nodes at all. 

\subsection{Experimental Setting}
We apply Algorithm~\ref{alg:detect} to the extracted Ethereum transaction graph $G=(V,E)$ to find instances of pass-through templates. 

We set the user-specified parameters in Conditions~II--IV of Section~\ref{sec:qualifying} to capture repeated rapid pass-through of substantial value. In Condition~II, we set $\bar{T}=1000$ blocks to require that value passes through the node within a short time window. At a roughly 12--13-second Ethereum block time, this corresponds to about 3.3--3.6 hours. In Condition~III, we set $\epsilon_{\mathrm{sim}}=0.99$ so that matched amounts must be nearly equal, while still allowing small differences due to fees or rounding. In Condition~IV, we set $p_{\mathrm{min}}=5000$ to exclude small transfers. We require that the matched pass-through template contains at least $m_{\min}=3$ matched pairs, that is, for a pass-through node $x_c$ we require $|M^\star(x_c)| \ge m_{\min}$. This ensures repeated behavior. These parameter choices are conservative and are intended to highlight only clearly suspicious activity.

\section{Results}
\label{sec:results}
This section reports the detected pass-through activity, the resulting layering network, and the broader structural properties and value flows of its nodes in the analyzed Ethereum transaction graph.
\subsection{Layering Network Identification}
\label{sec:raw-output}
Applying Algorithm~\ref{alg:detect} to the Ethereum transaction graph $G=(V,E,\lambda_E)$, we identify \textbf{283 pass-through template matches}. These are subgraphs $\{G_{t}^{\star}(x_i) = (V_{t}^{\star}(x_i), E_{t}^{\star}(x_i))\}_{i=1}^{283}$ of $G$, each centered on a distinct pass-through node $x_i \in V$ and having the structure described in Section~\ref{sec:qualifying}. The matching procedure over all nodes in $V$ takes less than two seconds, on a workstation with two AMD EPYC 9575F processors (128 cores, 256 threads total).

Fig.~\ref{fig:fullgraph} illustrates in red the identified pass-through nodes in the context of the Ethereum network. 

We call the graph formed by the union of all identified pass-through subgraphs the \emph{Layering Network}. Formally, the layering network is the graph
\begin{equation}
  \label{eq:layering-network-definition}
  G_{\mathrm{layer}} = \left(\bigcup_{i=1}^{n_\star} V_{t}^{\star}(x_i), \bigcup_{i=1}^{n_\star} E_{t}^{\star}(x_i), \lambda_E\right),
\end{equation}
where $n_\star:=283$ is the total number of pass-through template matches. We define 

\begin{equation}
\label{eq:layering-network-nodes-edges}
V_{\mathrm{layer}} := \bigcup_{i=1}^{n_\star} V_{t}^{\star}(x_i) \quad \text{ and } \quad E_{\mathrm{layer}} := \bigcup_{i=1}^{n_\star} E_{t}^{\star}(x_i),
\end{equation}
 as the sets of nodes and edges, respectively. Statistics about the output of Algorithm~\ref{alg:detect} and the layering network are summarized in Table~\ref{tab:final-summary}.

\begin{table}[t]
\centering
\caption{Output of Algorithm~\ref{alg:detect} applied to the Ethereum transaction graph $G$, and the graphs derived from it.}
\label{tab:final-summary}
\setlength{\tabcolsep}{1pt}
\begin{tabular}{@{}lrr@{}}
\toprule
\multicolumn{3}{@{}l}{\textit{Matching output}} \\
 & \multicolumn{2}{r}{\textbf{Value}} \\
\midrule
\multicolumn{2}{@{}l}{Number of template matches ($\{G_t^{\star}(x_i)\}$, \eqref{eq:final-matched-template-graph_1})} & 283 \\
\multicolumn{2}{@{}l}{Total distinct nodes matched ($|V_{\mathrm{layer}}|$, \eqref{eq:layering-network-nodes-edges})} & 1{,}690 \\
\multicolumn{2}{@{}l}{Matched transaction pairs ($|\mathcal{M}|$, \eqref{eq:matched-pairs_union})} & 3{,}510 \\
\midrule
\multicolumn{3}{@{}l}{\textit{Graphs built from}} \\
 \textit{the matching output} & \textbf{Nodes} & \textbf{Edges} \\
\midrule
$G_{{\scriptscriptstyle \mathrm{layer}}}= (V_{\scriptscriptstyle\mathrm{layer}}, E_{\scriptscriptstyle\mathrm{layer}})$ (layering network, \eqref{eq:layering-network-definition}) & 1{,}690 & 6{,}781 \\
$G_{\scriptscriptstyle\mathrm{induced}} = (V_{\scriptscriptstyle\mathrm{layer}}, E_{\scriptscriptstyle\mathrm{induced}})$ (induced subnetwork, \eqref{eq:induced-subgraph-edges}) & 1{,}690 & 20{,}541 \\
\bottomrule
\end{tabular}
\end{table}
The layering network splits into six connected components. One of them holds 98.3\% of all identified nodes. Thus, nearly all matched pass-through templates form a single connected structure rather than separate subgraphs, consistent with an organized layering structure. 
The OFAC-sanctioned seed address lies in the largest connected component of the layering network. Cross-referencing the other nodes in the layering network with the OFAC sanctions list reveals five additional sanctioned addresses in the same component. Together, these six addresses provide evidence linking the identified network to criminal entities designated by OFAC.

\begin{figure*}[t]
\centering
\includegraphics[width=0.9\textwidth]{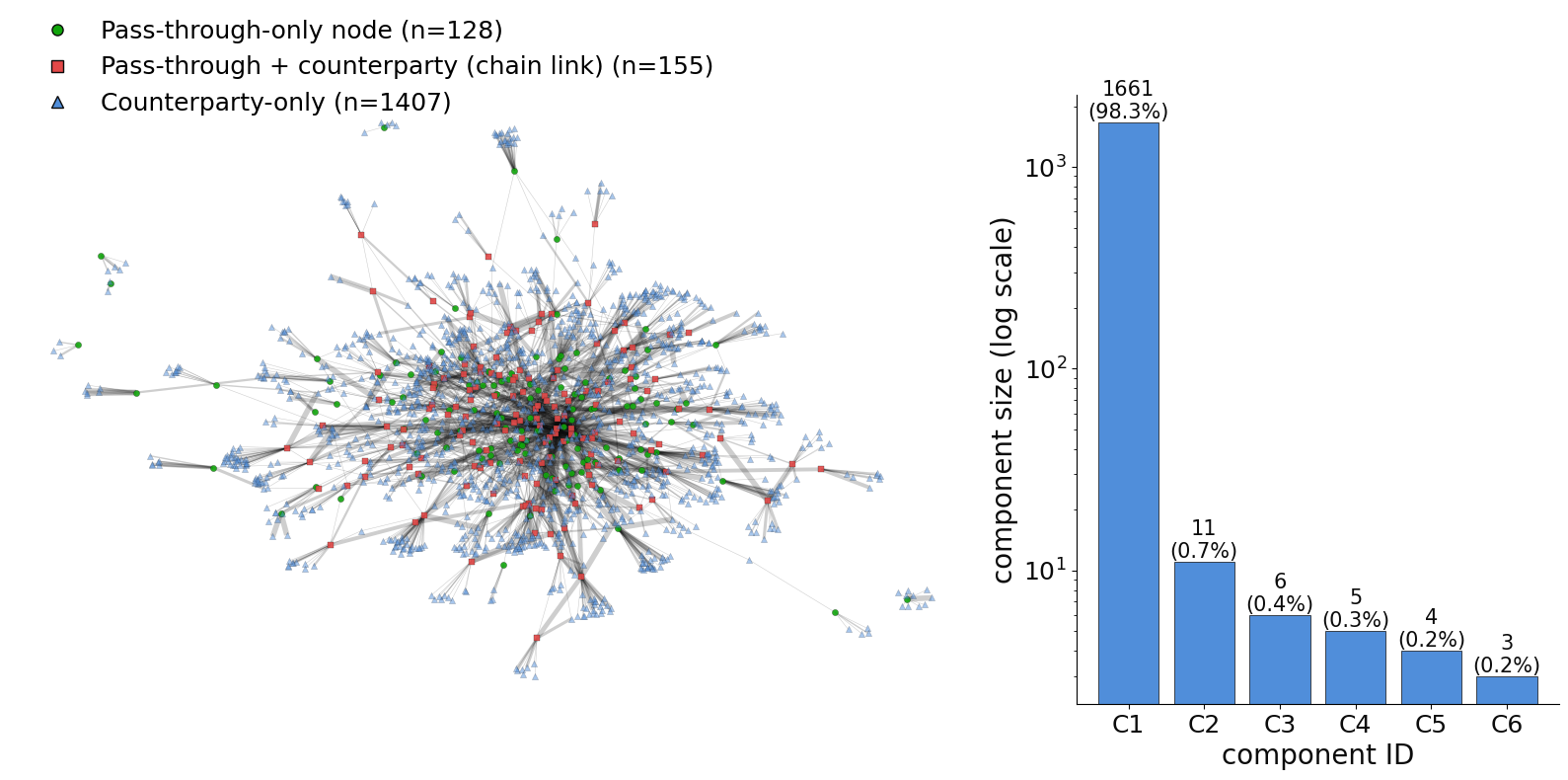}
\caption{\emph{Left:} The layering network $G_{\mathrm{layer}}$ (1{,}690 nodes). Nodes are colored by role. Pass-through-only nodes are flagged pass-through nodes with no other role in the network. Chain-link nodes are pass-through nodes that are also a counterparty of another pass-through node, linking two templates together. Counterparty-only nodes appear only as a sender or receiver in a matched pair, never as a pass-through node themselves. Multiedges are collapsed into a single edge for clarity. \emph{Right:} Number of nodes in each of the six connected components of $G_{\mathrm{layer}}$, on a logarithmic scale. The largest component holds 98.3\% of all nodes.}
\label{fig:layering_network}
\end{figure*}

Fig.~\ref{fig:layering_network} displays the layering network, with nodes colored according to their role, together with the component-size distribution on a logarithmic scale.

\subsection{Analyzing Value Flow in the Layering Network}
\label{sec:analyzing-value-flow-in-layering-network}
We measure the total value transferred through $G_{\mathrm{layer}}$, which is built from matched pass-through templates $G_t^{\star}(x_i) = (V_t^{\star}(x_i), E_t^{\star}(x_i))$. Each matched template is defined by a set of edge pairs $M^{\star}(x_i)$ (see \eqref{eq:matched-qualifying-pairs} and \eqref{eq:final-matched-template-edges}), where each pair consists of an incoming and outgoing edge at the pass-through node $x_i$, representing funds received and then forwarded. Summing the value of every edge in $G_{\mathrm{layer}}$ directly would double count these funds, once as received and once as forwarded, so instead we collect the matched pairs $M^{\star}(x_i)$ across all pass-through nodes and define
\begin{equation}
  \label{eq:matched-pairs_union}
\mathcal{M} := \bigcup_{i=1}^{n_\star} M^{\star}(x_i).
\end{equation}
We then define the total per-hop transferred value by
\begin{equation}
\label{eq:per-hop-value}
F_{\mathrm{hop}}
:=
\sum_{(e^{\mathrm{in}},e^{\mathrm{out}})\in \mathcal{M}}
\min\{p_{e^{\mathrm{in}}},\, p_{e^{\mathrm{out}}}\},
\end{equation}
where $p_{e^{\mathrm{in}}}$ and $p_{e^{\mathrm{out}}}$ are the amounts transacted on the incoming and outgoing edges, respectively. We apply formula \eqref{eq:per-hop-value} to the layering network. The total transacted value, \(F_{\mathrm{hop}}\), is \textbf{\$286.4 million} (Table~\ref{tab:value-moved}, ``per-hop''). Note that we compute a conservative estimate: for each matched pair we only consider the smaller of the incoming and outgoing amounts, that is, $\min\{p_{e^{\mathrm{in}}},p_{e^{\mathrm{out}}}\}$. The incoming and outgoing amounts are numerically identical ($p_{e^{\mathrm{in}}} = p_{e^{\mathrm{out}}}$) for 2{,}604 of identified matched pairs (74.19\%), while the remaining 906 (25.81\%) satisfy the $99\%$ similarity tolerance without being exactly equal.
The identified layering network, \(G_{\mathrm{layer}}\), transacts four USD-pegged stablecoins~\cite{watsky2024stablecoins}: USDT, USDC, BUSD, and DAI. We therefore treat one unit of any of these assets as one U.S. dollar and aggregate values across assets without currency conversion.
\begin{table}[t]
\centering
\caption{Total USD value moved in the identified layering network, $G_{\mathrm{layer}}$, by asset. Values are reported in USD millions.}
\label{tab:value-moved}
\begin{tabular}{@{}lrr@{}}
\toprule
& \multicolumn{2}{c}{\textbf{USD value (millions)}} \\
\cmidrule(l){2-3}
\textbf{Asset} & \textbf{Min per hop} & \textbf{Chain-adjusted} \\
\midrule
BUSD & 1.49   & 1.49   \\
DAI  & 0.23   & 0.23   \\
USDC & 44.54  & 43.15  \\
USDT & 240.14 & 218.83 \\
\midrule
\textbf{ALL (pooled)} & \textbf{286.40} & \textbf{263.70} \\
\bottomrule
\end{tabular}
\end{table}

$F_{\mathrm{hop}}$ may double count value because a single transfer edge can belong to two distinct matched pass-through templates, once as an outgoing edge and once as an incoming edge. This occurs when the same funds pass through multiple nodes. To avoid this double counting, we group the matched pairs in $\mathcal{M}$ into chains. A chain is a maximal sequence of pairs linked end to end, such that the outgoing edge of each pair is the incoming edge of the next. Let $\mathcal{C}_1,\ldots,\mathcal{C}_r$ denote the resulting chains, and let $k_l := |\mathcal{C}_l|$ denote the length of chain $\mathcal{C}_l$, so that $ \mathcal{M} = \bigcup_{l=1}^{r} \mathcal{C}_l$. A chain with $k_l=1$ consists of a single matched pair, as in $F_{\mathrm{hop}}$. Appendix~\ref{app:chains} gives the full index-level definition. The total de-duplicated value is then
\[
F_{\mathrm{chain}} := \sum_{l=1}^{r} \min_{(e^{\mathrm{in}},\, e^{\mathrm{out}}) \,\in\, \mathcal{C}_l} \min\{p_{e^{\mathrm{in}}}, p_{e^{\mathrm{out}}}\}.
\]
This gives
\(F_{\mathrm{chain}}=\textbf{\$263.7 million}\)
(Table~\ref{tab:value-moved}, ``chain-adjusted''). There are 3{,}051 single-hop chains ($k_l=1$), 203 two-hop chains ($k_l=2$), 15 three-hop chains ($k_l=3$), and 2 four-hop chains ($k_l=4$).  

Table~\ref{tab:distribution} reports the distribution of individual transfer amounts among the 6{,}781 distinct directed edges retained in the final 1{,}690-node layering network $G_{\mathrm{layer}}$.

We remark that $F_{\mathrm{hop}}$ and $F_{\mathrm{chain}}$ quantify only the value flow captured by the layering network $G_{\mathrm{layer}} = (V_{\mathrm{layer}}, E_{\mathrm{layer}})$, which represents only pass-through transaction activity. They do not measure the full
transaction activity of nodes in $V_{\mathrm{layer}}$ in the complete Ethereum transaction graph analyzed in this work. 
\begin{table}[t]
\centering
\caption{Number of transfers and USD amount summary by asset over the directed edges of the identified layering network $G_{\mathrm{layer}}$. Number of transfers is reported exactly. All USD amount columns are rounded to the nearest thousand and reported in thousands of USD (e.g., 5 = \$5{,}000).}
\label{tab:distribution}
\begin{tabular}{@{}lrrrrr@{}}
\toprule
& & \multicolumn{4}{c}{\textbf{USD value (thousands)}} \\
\cmidrule(l){3-6}
\textbf{Asset} & \shortstack{\textbf{Number of} \\ \textbf{transfers}} & \textbf{Min} & \textbf{Max} & \textbf{Median} & \textbf{Mean} \\
\midrule
USDT  & 6{,}030 & 5  & 6{,}800 & 42 & 76 \\
USDC  & 719     & 5  & 1{,}963 & 49 & 122 \\
BUSD  & 20      & 17 & 462     & 71 & 149 \\
DAI   & 12      & 10 & 90      & 21 & 38 \\
\bottomrule
\end{tabular}
\end{table}
\subsection{Broader Activity of Layering Network Nodes in the Ethereum Transaction Graph}
\label{sec:broader-activity-of-layering}
We now examine the full activity of the layering network nodes within the Ethereum graph $G$. The layering network $G_{\mathrm{layer}} = (V_{\mathrm{layer}}, E_{\mathrm{layer}})$ contains only the matched pass-through edges, so it captures a narrow slice of what its nodes actually do: in the full graph $G$, these same nodes send and receive many other transactions. To recover that broader activity, let $G_{\mathrm{induced}} = (V_{\mathrm{layer}}, E_{\mathrm{induced}})$ be the subgraph of $G$ induced by $V_{\mathrm{layer}}$, where
\begin{equation}
\label{eq:induced-subgraph-edges}
E_{\mathrm{induced}} := \{e \in E : \mathrm{src}(e), \mathrm{dst}(e) \in V_{\mathrm{layer}}\}
\end{equation}
is the set of every edge of $G$ with both endpoints in $V_{\mathrm{layer}}$. There are 20{,}541 such edges in total. The rest of this section compares the connectivity of $G_{\mathrm{induced}}$ to that of $G$, then examines how value flows between the nodes in the layering network and the rest of $G$.

\subsubsection{Connectivity Analysis} 
We compare how connected $G_{\mathrm{induced}}$ is, relative to the full Ethereum network $G$, by considering their simple undirected projections $\Pi(G)$ and $\Pi(G_{\mathrm{induced}})$. A simple undirected projection of a transaction graph replaces directed edges with a single undirected edge between two nodes whenever at least one edge occurs in either direction. Appendix~\ref{app:topology-definitions} gives the formal definition of projection.

\begin{figure*}[t]
\centering
\includegraphics[width=0.95\linewidth]{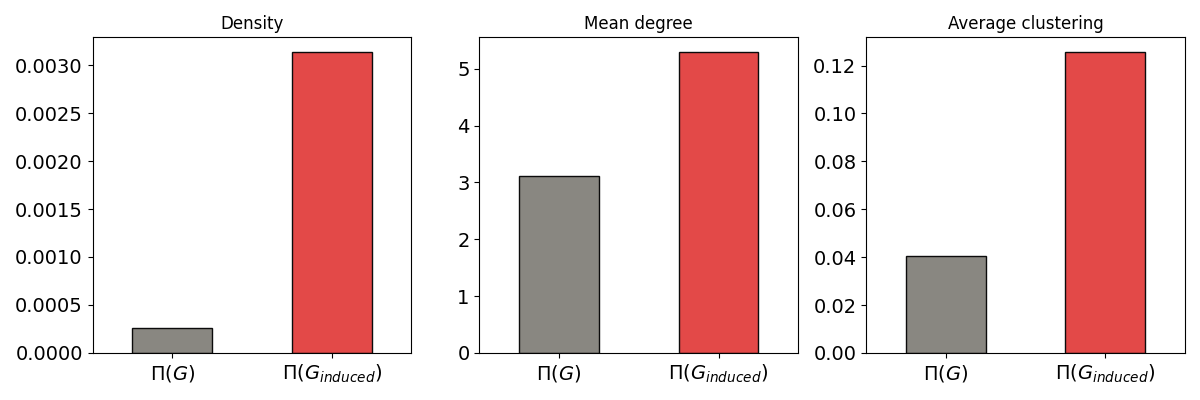}
\caption{Density, mean degree, and average clustering coefficient of \(\Pi(G_{\mathrm{induced}})\), the simple undirected projection of the Ethereum subgraph induced by the nodes in the layering network, compared to \(\Pi(G)\), the projection of the full Ethereum graph.}
\label{fig:topology}
\end{figure*}

Fig.~\ref{fig:topology} reports metrics comparing the topology of the projection of $G_{\mathrm{induced}}$ against that of $G$. In particular, we consider density, mean degree and average clustering, as formally defined in Appendix~\ref{app:topology-definitions}. $G_{\mathrm{induced}}$ is an order of magnitude denser (density 0.0031 vs.\ 0.0003), has a higher mean degree (5.29 vs.\ 3.11), and is more than three times as clustered (average clustering 0.1255 vs.\ 0.0406). This gap is consistent with the layering network being an organized structure rather than an arbitrary subset of nodes. 

\begin{figure}[t]
\centering
\includegraphics[width=0.95\linewidth]{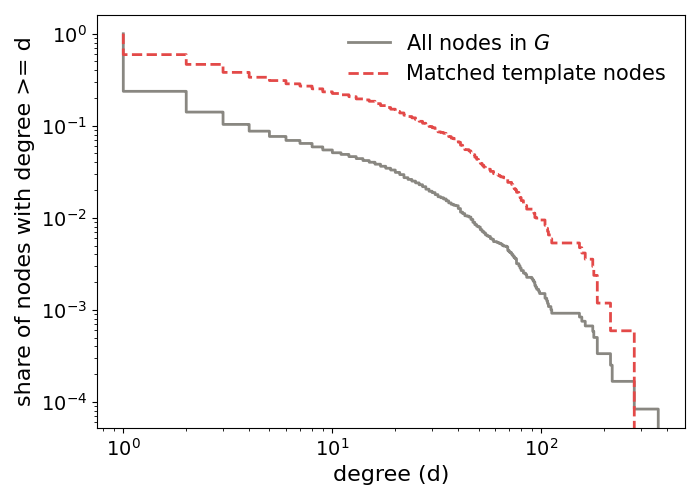}
\caption{Cumulative degree distribution (share of nodes with degree $\geq d$ in $\Pi(G)$) on log-log axes, comparing all nodes of the full Ethereum network $G$ to the subset of nodes in $V_{\mathrm{layer}}$, with degree measured using the edge set of $\Pi(G)$ for both. The $V_{\mathrm{layer}}$ curve lies above the all-node curve across nearly the full range.}
\label{fig:degree-ccdf}
\end{figure}

We also compute the degree distribution of the nodes in $V_{\mathrm{layer}}$ within the projection of the full Ethereum network $\Pi(G)$, rather than restricting to their induced subgraph $\Pi(G_{\mathrm{induced}})$. Fig.~\ref{fig:degree-ccdf} compares, for each degree threshold $d$, the share of nodes whose degree in $\Pi(G)$ is at least $d$, computed once over all nodes of $\Pi(G)$ and once restricted to the nodes in $V_{\mathrm{layer}}$. The $V_{\mathrm{layer}}$ curve lies above the all-node curve across nearly the full range, showing that these nodes are consistently more highly connected within the full network, not only to each other.
\subsubsection{Value Flow Analysis}
We analyze how value flows through the nodes of the identified layering network within the full Ethereum graph $G=(V,E)$. We restrict the analysis to stablecoin transfers (USDT, USDC, BUSD, DAI, all pegged to one US dollar).

As described in Section~\ref{sec:building-the-ethereum-transaction-graph}, the Ethereum graph $G$ contains a set of boundary nodes $H \subset V$. These nodes may act as gateways, injecting or absorbing value from outside the network, as exchange wallets do when aggregating deposits and withdrawals from unrelated senders and receivers. For simplicity, we treat all boundary nodes as potential gateways. Within the layering network, let $H_{\mathrm{layer}} = V_{\mathrm{layer}} \cap H$ denote its boundary nodes and define the core as $V_{\mathrm{core}} = V_{\mathrm{layer}} \setminus H_{\mathrm{layer}}$, with $|H_{\mathrm{layer}}| = 66$ and $|V_{\mathrm{core}}| = 1{,}624$.  This split allows us to distinguish value circulating within the core from value entering or leaving the layering network through boundary nodes. Of the 66 boundary nodes, 63 appear only as senders or receivers in the layering network, consistent with their role as gateways to the broader Ethereum ecosystem rather than as drivers of pass-through behavior.

Recall that for edge $e\in E$ its label is $\lambda_E(e) = (a_e, p_e, \tau_e)$, where $a_e$ and $p_e$ are the asset and amount transferred on edge $e$. We define the stablecoin edges of $G$ as
$$E_{\$} := \{e \in E : a_{e} \in \{\mathrm{USDT}, \mathrm{USDC}, \mathrm{BUSD}, \mathrm{DAI}\}\}.$$ We now quantify how much stablecoin value flows into, out of, and within the core $V_{\mathrm{core}}$, relative to the rest of $G$. For \(A, B \subseteq V\), let
\[
F(A,B) := \sum_{\substack{e \in E_{\$}, \\ \mathrm{src}(e) \in A, \mathrm{dst}(e) \in B}} p_{e}.
\]
Then
\[
F_{\mathrm{gross}} := F(V_{\mathrm{core}}, V_{\mathrm{core}}), \qquad
F_{\mathrm{in}} := F(V \setminus V_{\mathrm{core}}, V_{\mathrm{core}}),
\]
\[
F_{\mathrm{out}} := F(V_{\mathrm{core}}, V \setminus V_{\mathrm{core}}), \qquad
R:=\frac{F_{\mathrm{gross}}}{F_{\mathrm{in}}},
\]
where $V \setminus V_{\mathrm{core}}=\{v\in V : v \notin V_{\mathrm{core}}\}$.

\(F_{\mathrm{gross}}\) captures internal circulation within the core $V_{\mathrm{core}}$, \(F_{\mathrm{in}}\) captures inflow from the boundary nodes or outside nodes, and \(F_{\mathrm{out}}\) captures outflow from the core. The reuse ratio \(R\) measures how many times value entering the core is retransferred within it.
Pooled across assets, the core moves
\(F_{\mathrm{gross}} \approx \$1.04\) billion internally, receives
\(F_{\mathrm{in}} \approx \$410\) million from outside, and sends out
\(F_{\mathrm{out}} \approx \$379\) million. The pooled multiplier is $R=2.53$: each dollar that enters the core is retransferred within it about two and a half times.
Table~\ref{tab:flow-reuse} reports the computed quantities for the four USD-pegged stablecoins.
\begin{table}[t]
\centering
     \caption{Directed stablecoin value-flow decomposition for the core node set, broken down by asset. $F_{\mathrm{gross}}$ quantifies transfers between core nodes. $F_{\mathrm{in}}$ quantifies inflow transfers from boundary nodes or from nodes not in $V_{\mathrm{layer}}$ into the core. $F_{\mathrm{out}}$ quantifies outflow transfers from the core to boundary nodes or to nodes outside $V_{\mathrm{layer}}$. Values are reported in USD millions.}
\label{tab:flow-reuse}
\begin{tabular}{@{}lrrrr@{}}
\toprule
& \multicolumn{3}{c}{\textbf{USD value (millions)}} & \\
\cmidrule(l){2-4}
\textbf{Asset} & \textbf{$F_{\mathrm{gross}}$} & \textbf{$F_{\mathrm{in}}$} & \textbf{$F_{\mathrm{out}}$} & \textbf{$R$} \\
\midrule
BUSD & 2.84 & 1.38 & 0.16 & 2.06 \\
DAI & 0.16 & 0.24 & 0.33 & 0.67 \\
USDC & 160.99 & 23.93 & 21.78 & 6.73 \\
USDT & 871.74 & 384.08 & 356.64 & 2.27 \\
\midrule
\textbf{All (pooled)} & \textbf{1{,}035.73} & \textbf{409.63} & \textbf{378.91} & \textbf{2.53} \\
\bottomrule
\end{tabular}
\end{table}
\subsection{Node Types in the Layering Network}
Blockchain address nodes can be either externally owned accounts (EOAs), which are controlled by private keys, or smart contracts, which execute code when called. Distinguishing between them helps separate human-controlled activity from automated protocol interactions in the layering network. Using the Alchemy API~\cite{alchemy_ethereum_api}, we classify all 1{,}690 addresses in the identified layering network and find 39 smart contracts. Only three of these contracts act as pass-through nodes (two pass-through-only and one chain-link), while the remaining 36 appear only as counterparties, including 28 receivers, 13 senders, and 5 addresses in both roles. We retain these contract counterparties because they may represent legitimate parts of a real layering flow, such as decentralized exchange routers, aggregators, or intermediary pool contracts. Only three of the 283 pass-through nodes (1.06\%) are contracts. The repeated pass-through pattern is driven overwhelmingly by EOAs, not automated contracts. 

\section{Conclusion}
\label{sec:conclusion}
We develop a framework for detecting pass-through patterns consistent with layering and apply it to Ethereum. Algorithm~\ref{alg:detect} identifies 283 matches of pass-through templates involving 1{,}690 address nodes, 98.3\% of which lie in a single connected component containing six OFAC-sanctioned addresses. Matched pass-through transfers move over \$260 million. The full activity of the matched address nodes forms a subnetwork markedly denser than an average slice of the analyzed Ethereum network. A core set of matched nodes moves approximately \$1.04 billion in stablecoins, with \$410 million entering and \$379 million leaving the core. Each dollar entering the core is retransferred internally 2.53 times on average.

These results show that a rule-based, interpretable approach can uncover large, structured transaction subnetworks without model training or opaque anomaly scores. Independent of this case study, the method offers three properties that generalize beyond it. First, every match is auditable because pass-through templates are defined by explicit criteria on asset, timing, and amount (Conditions~I--IV). An analyst can check any matched instance directly against those criteria. Second, matching is solved exactly rather than heuristically: Edmonds' blossom algorithm provably returns the maximum-cardinality, maximum-weight optimal set of pass-through pairs at each node (Section~\ref{sec:optimization}). Third, the procedure is efficient. It runs on the full 11,978-node, 49,921-edge Ethereum graph in under two seconds, with cost that grows linearly in the number of nodes once per-node degree is bounded (Section~\ref{sec:optimization}). Because the transaction-graph formalism of Section~\ref{sec:blockchain-graph} places no assumptions on the underlying network beyond directed edges labeled with an asset, an amount, and a timestamp, the same procedure applies unchanged to transaction data from other blockchains, or to any setting with labeled transaction records, not only Ethereum.

\section*{Ethical and Legal Considerations}
\label{sec:ethics}

All data underlying this study are public: the Ethereum transaction data are drawn from the AWS Public Blockchain Data program~\cite{aws_public_blockchain}, and the sanctions list is publicly released by the U.S. Department of the Treasury~\cite{ofac_sdn_list}. Ethereum addresses are pseudonymous identifiers.



\appendices
\section{Chain Decomposition}
\label{app:chains}
This appendix gives the full definition of the chain decomposition used in Section~\ref{sec:analyzing-value-flow-in-layering-network} to compute $F_{\mathrm{chain}}$.

Recall that \(\mathcal{M}\) is the set of all matched incoming/outgoing edge pairs across all pass-through nodes (see \eqref{eq:matched-pairs_union}). We partition \(\mathcal{M}\) into linked components by following an outgoing edge of one pair to the incoming edge of the next. Because each pair in \(\mathcal{M}\) has at most one predecessor and at most one successor, each component is either a simple path or a simple cycle. We denote the \(l\)th component by
\[
\mathcal{C}_l=\{(e_{1,l}^{\mathrm{in}},e_{1,l}^{\mathrm{out}}),\ldots,(e_{k_l,l}^{\mathrm{in}},e_{k_l,l}^{\mathrm{out}})\},
\]
where \((e_{j,l}^{\mathrm{in}},e_{j,l}^{\mathrm{out}})\in\mathcal{M}\) is the \(j\)th pair of the \(l\)th component, \(k_l\) is its length, and
\[
e_{j,l}^{\mathrm{out}}=e_{j+1,l}^{\mathrm{in}}, \qquad j=1,\ldots,k_l-1.
\]
In the cyclic case, we additionally have \(e_{k_l,l}^{\mathrm{out}}=e_{1,l}^{\mathrm{in}}\). Thus each \(\mathcal{C}_l\) is a maximal linked component of \(\mathcal{M}\). If \(k_l=1\), it reduces to a single matched pair, exactly as in \(F_{\mathrm{hop}}\) in \eqref{eq:per-hop-value}. Assuming there are \(r\) components in total, \(\mathcal{M}=\bigcup_{l=1}^r \mathcal{C}_l\), and every pair in \(\mathcal{M}\) belongs to exactly one \(\mathcal{C}_l\).

\section{Topology Definitions}
\label{app:topology-definitions}

For a directed multigraph $G=(V,E)$, its simple undirected projection keeps the same vertex set \(V\) and replaces \(E\) by

\begin{align*}
E^{\mathrm{u}} := \bigl\{&(u,v) \in V\times V :\ \exists e\in E \\
&\text{s.t. } (\mathrm{src}(e),\mathrm{dst}(e)) \in \{(u,v),(v,u)\}\bigr\}.
\end{align*}

Note that 
$E^{\mathrm{u}} \subseteq V \times V, \;\text{with } (u,v) \equiv  (v,u)$. Thus all directions and multiplicities are discarded. The projection of $G$ is defined as $$\Pi(G):=(V, E^{\mathrm{u}}).$$
For a simple undirected graph with vertex set \(V\) and edge set \(E^{\mathrm{u}}\), write \(n:=|V|\), \(m:=|E^{\mathrm{u}}|\), and \(d(v)\) for the degree of \(v\in V\). The density is
$
\mathrm{dens}(V,E^{\mathrm{u}}):=\frac{2m}{n(n-1)}.
$
The mean degree is
$
\overline{d}(V,E^{\mathrm{u}}):=\frac{1}{n}\sum_{v\in V} d(v)=\frac{2m}{n}.
$

For each vertex \(v\in V\), let \(N(v)\) be its 1-hop neighborhood and let
$
t(v):=
\bigl|\bigl\{(u,w)\in E^{\mathrm{u}} : u,w\in N(v)\bigr\}\bigr|,
$
be the number of edges among its neighbors. The local clustering coefficient is
\[
c(v):=
\begin{cases}
\dfrac{2\,t(v)}{d(v)\bigl(d(v)-1\bigr)}, & d(v)\ge 2,\\[6pt]
0, & d(v)<2,
\end{cases}
\]
and the average clustering coefficient is
$
\mathrm{clust}(V,E^{\mathrm{u}}):=\frac{1}{n}\sum_{v\in V} c(v).
$

\section{Digital Assets Considered}
\label{app:stablecoins}
Stablecoins are crypto-assets designed to maintain a stable value relative to a noncrypto reference asset, most commonly the U.S. dollar~\cite{watsky2024stablecoins}. We consider four U.S.-dollar stablecoins: USDT (Tether), USDC (USD Coin), and BUSD (Binance USD), which are backed by off-chain reserve assets, and DAI, which maintains its dollar peg primarily through crypto-asset collateral and decentralized protocols. We also include ETH, the native asset of Ethereum, and WETH (Wrapped Ether), an ERC-20 representation of ETH that is designed to track its value one-to-one.

\bibliographystyle{IEEEtran}
\bibliography{IEE_references}
\end{document}